\documentclass[twocolumn,showpacs,preprintnumbers,amsmath,amssymb]{revtex4}

\usepackage{graphicx}
\usepackage{dcolumn}
\usepackage{bm}
\usepackage{epsfig}

\begin{document}

\title{Surface freezing and a two-step pathway of the isotropic-smectic phase transition in colloidal rods}

\author{Zvonimir Dogic}
\affiliation{IFF/Weiche Materie, Forschungszentrum J\"{u}lich,
J\"{u}lich, D-52425 Germany}


\date{\today}

\begin{abstract}
We study the kinetics of the isotropic-smectic phase transition in
a colloidal rod/polymer mixture by visualizing individual smectic
layers. First, we show that the bulk isotropic-smectic phase
transition is preceded by a surface freezing transition in which a
quasi two-dimensional smectic phase wets the isotropic-nematic
interface. Next, we identify a two step kinetic pathway for the
formation of a bulk smectic phase. In the first step a metastable
isotropic-nematic interface is formed. This interface is wetted by
the surface induced smectic phase. In the subsequent step, smectic
layers nucleate at this surface phase and grow into the isotropic
bulk phase.
\end{abstract}


\maketitle

Colloids with hard core repulsive interactions are often studied
due to the simplicity and generality of their intermolecular
potential. As a result of these studies, the equilibrium phase
diagram of hard rods and spheres is well understood at the present
time~\cite{Frenkel88a,Vroege92,Pusey86,Adams98}. However, much
less is known about the kinetic pathways of phase transitions in
these systems~\cite{tenWolde97,Zhu97}. Direct visualization of
colloids in a system undergoing phase transition have provided a
powerful tool to study general aspects of phase transition
kinetics~\cite{Gasser01,Anderson02}. In this paper we study the
kinetics of the isotropic-smectic phase transition by directly
visualizing individual smectic layers in a phase separating
sample. As a model system of colloidal rods we use a monodisperse
suspension of {\it fd} virus~\cite{Dogic01}. We elucidate a
kinetic pathway of unexpected complexity. The existence of surface
freezing and a metastable isotropic-cholesteric phase transitions
is discovered and their influence on the kinetic pathway is
discussed. Because the behavior of the {\it fd}/Dextran mixture is
determined by steric interactions and since all molecules
including low molecular weight thermotropics have a steric core
the results reported in this paper are likely to be quite general.
In addition, our results might be pertinent to understanding  the
dynamics of amphiphilic membranes~\cite{Lipowsky95}, 2D smectic
systems~\cite{Harrison00}, surface freezing and wetting
transitions~\cite{Wu93b,Ocko86,Bonn01} and self assembled
nano-structures~\cite{Lee02}.

It has been known for a long time that surface freezing/melting
can dramatically alter the nucleation rate and the kinetic pathway
of a phase transition. On one hand, most substances exhibit
surface melting. In this case a liquid surface wets the
crystalline bulk phase. It follows that crystals melt from the
surface inwards and therefore it is difficult to prepare a
superheated metastable solid~\cite{Veen99,Cahn86}. On the other
hand, surface freezing is observed in very few systems, most
notably thermotropic liquid crystals, alkanes and surfactant
mesophases~\cite{Wu93b,Ocko86,Lang99}. Upon supercooling these
materials, the ordered phase nucleates at the frozen interface and
propagates towards the bulk phase. Therefore, it is difficult to
supercool liquids that exhibit surface
freezing~\cite{Sloutskin01}.

\begin{figure}
\centerline{\epsfig{file=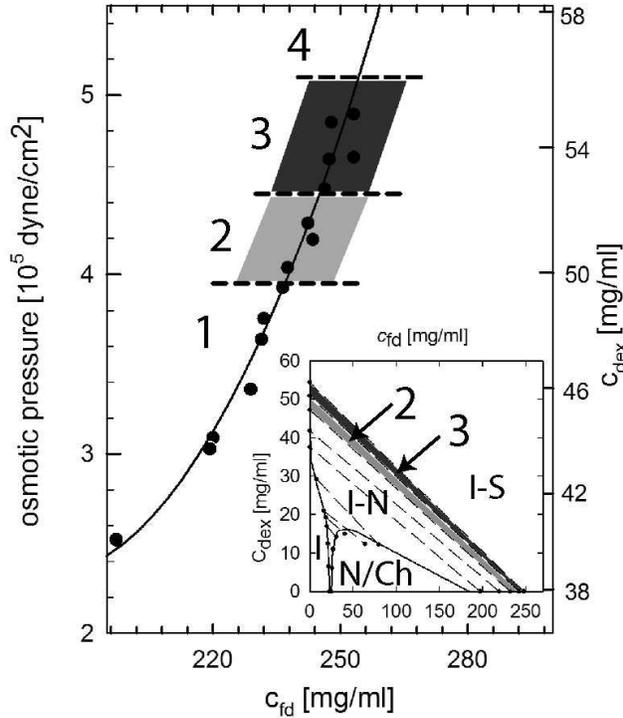,width=8.5cm}}
\caption{\label{figure1} The coexistence concentrations of an
immiscible {\it fd}-Dextran mixture. The Y-axes shows
concentration and osmotic pressure of Dextran in the phase that
coexists with rod-rich liquid crystalline phase(droplets) whose
concentration is shown on the X-axis. Numbers 1 through 4 indicate
regions where different phase behaviors are observed. Images of
the structures observed in these regions are shown in
Fig.~\ref{figure2}. Stable surface smectic phase wets the
isotropic-nematic interface in region 2. Colloidal membranes are
stable in region 3. Inset: The complete phase diagram of {\it
fd}/Dextran mixture. Tie lines along which the phase separation
proceeds are indicated by dashed lines. Regions of the
isotropic-nematic (I-N) and isotropic-smectic (I-S) coexistence
are indicated. }
\end{figure}

Another factor that can affect the nucleation rate of a
transitions is the presence of metastable
phases~\cite{Sirota99,tenWolde97}. For example, recent simulations
predict that the free energy barrier for the formation of protein
crystals is greatly reduced when a metastable gas-liquid phase
transition is located in a vicinity of a stable liquid-solid phase
boundary~\cite{tenWolde97}. In this case the nucleation of protein
crystals proceeds in two steps. In the first step a dense
metastable droplet associated with the gas-liquid phase transition
is formed, while in the subsequent step the protein crystal
nucleates within this droplet. In this paper we show that both
surface freezing and metastable phases are important for
understanding the kinetics of the isotropic-smectic phase
transition.

\begin{figure}
\centerline{\epsfig{file=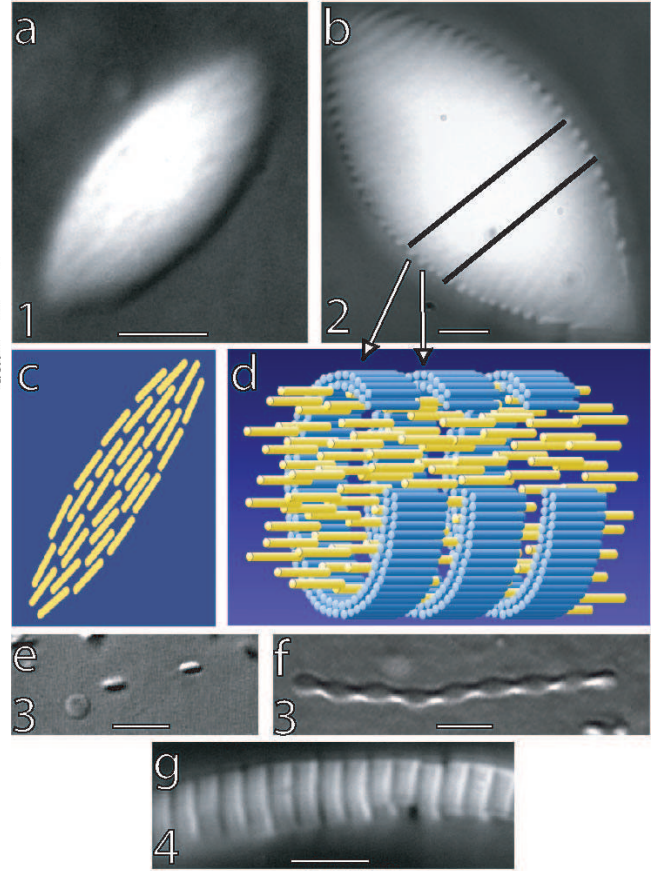,width=8.5cm}}
\caption{\label{figure2} Images and schematic representations of
different structures observed in the {\it fd}/Dextran mixture.
{\bf (a)} Anisotropic nematic droplet in the polymer rich
background. Configuration of rods in shown in figure c.  {\bf (b)}
Nematic droplet with surface smectic phase. Image is formed by
focusing on the midplane of the tactoids. The 3D structure is an
object of revolution about the long axis. A schematic
representation of a nematic droplet with surface induced smectic
phase is shown in image d. {\bf (e)} Colloidal membranes which
homogeneously nucleate from isotropic phase. The bottom left disk
lies in the plane of the image while the other two line
perpendicular to the plane of the image {\bf (f)} Twisted ribbon
which is identical to colloidal membrane except that that it is
elongated along the twist directions {\bf (g)} Bulk
isotropic-smectic phase coexistence. Scale bars indicate 3
$\mu$m.}
\end{figure}

Bacteriophage {\it fd} is a semi-flexible virus with contour
length of 880 nm, diameter of 7 nm and persistence length of
2200nm. It was prepared as previously described and dialyzed
against buffer of known ionic strength (190 mM Nacl,10 mM Tris, pH
= 8.10). The phase diagram of the rod-polymer mixture was measured
according to the published procedure~\cite{Dogic01}. All the
samples are prepared in a metastable/unstable isotropic phase by
shear melting any existing structure and samples are placed into
rectangular capillaries (VitroCom, Mountain Lakes, NJ). Nucleation
and growth of the order phase is observed with an optical
microscope (Zeiss AxioPlan2) equipped with DIC optics. All images
are recorded with a cooled CCD camera (AxioCam Zeiss)

At zero polymer concentration {\it fd} is a good model system of
hard rods and forms a stable isotropic (I), cholesteric (Ch) and
smectic (S) phases with increasing concentration in agreement with
theoretical predictions~\cite{Dogic97,Frenkel88a}. Equilibrium I-S
phase transition is observed in a mixture of rod-like {\it fd}
viruses and non-adsorbing polymer Dextran. The phase diagram of
this mixture is shown in the inset of Fig.~\ref{figure1}. Adding
non-adsorbing polymer to {\it fd} suspension produces effective
attractive interactions between {\it fd} rods~\cite{Asakura54}.
The main consequence of this attractive potential on the phase
behavior of a rod like system is to widen the I-Ch coexistence
concentrations with the polymer preferentially partitioning into
the isotropic phase~\cite{Bolhuis97}. Since the interactions in
the {\it fd}/polymer mixtures are temperature independent, all
phase transitions are entropically driven. In the first part of
the paper we describe the equilibrium structures related to the
surface freezing observed in region 2 of the phase diagram. In the
second part of the paper we describe one of the kinetic pathways
of phase separation observed in region 3.

\begin{figure}
\centerline{\epsfig{file=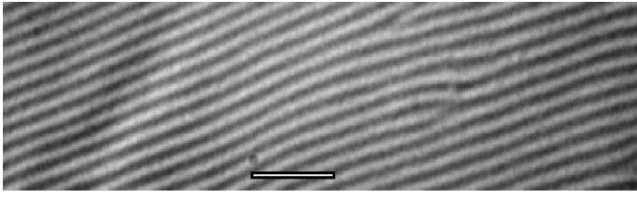,width=8.5cm}}
\caption{\label{figure3} Image of a macroscopically phase
separated isotropic-nematic interfaces which exhibit surface
freezing. The concentrations of the coexisting isotropic and
nematic phases are c$_{\mbox{\scriptsize {\it fd}}}=242$ mg/ml and
c$_{\mbox{\scriptsize dex}}=51.5$ mg/ml. Dense nematic phase is
below the image plane while the isotropic phase is above the
image. The thickness of the surface induced smectic phase is few
hundred nm. The surface structure show here is identical to the
surface of tactoids shown in Fig.~\ref{figure2}d. A pair of
dislocation defects is clearly visible in the image. Scale bars
indicate 5 $\mu$m.}
\end{figure}

At rod concentrations below 235 mg/ml (region 1 in
Fig.~\ref{figure1}a and b), nematic droplets (tactoids) form in an
isotropic background (Fig.~\ref{figure2}a). Polarization
microscopy indicates that the configuration of rods in the nematic
tactoid is as shown in Fig.~\ref{figure2}c. When confined to a
small volume the cholesteric order is not able to develop;
therefore we observe only unwound nematic phase within a
individual tactoid. At higher rod concentrations (region 2 in
Fig.~\ref{figure1}) we observe droplets that have the same
anisotropic shape. Microscopy indicates that the interior of these
droplets is still nematic. However, each droplet has a corrugated
I-N interface where the length of each ridge along the droplet's
long axis is approximately one virus long. As the tactoids
coalesce and increase in size, the surface corrugations are always
confined to a narrow layer of well defined thickness located at
the I-N interface. This implies that the formation of corrugations
is a purely surface effect. These observations lead us to
conclusion that there exists a surface-induced quasi 2D smectic
phase that wets the I-N interface. The ridges observed at the
interface are individual layers of the surface-induced smectic
phase. A schematic representation of a section of a corrugated
tactoid is shown in Fig.~\ref{figure2}d. The surface smectic phase
is observed above an {\it fd} concentration of 235 mg/ml while the
bulk I-S phase transition (region 4) is observed at 255 mg/ml.

\begin{figure}
\centerline{\epsfig{file=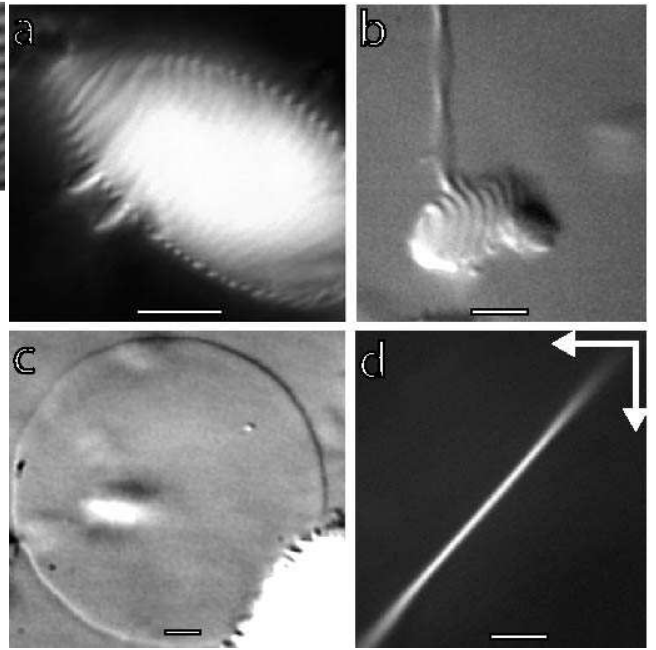,width=8.5cm}}
\caption{\label{figure4} Images of structures observed in region 3
of the phase diagram after the sample has been equilibrated for a
few days. For images a and b c$_{\mbox{\scriptsize {\it fd}}}$=254
mg/ml and c$_{\mbox{\scriptsize dex}}=53.5$ mg/ml while for images
c and d c$_{\mbox{\scriptsize dex}}=56$ mg/ml and
c$_{\mbox{\scriptsize {\it fd}}}$ is undetermined. {\bf (a)}
Nematic droplet with a surface frozen smectic phase. Surface
smectic phase acts as a nucleation site for the formation of
colloidal membranes. {\bf (b)} Twisted smectic ribbon nucleates at
the surface smectic phase and grows into the isotropic bulk phase
{\bf (c)} DIC image of a large (35 $\mu$m diameter) isolated
colloidal membrane in which rods lie perpendicular to the image
plane. Correspondingly the membrane shows no birefringence under
crossed polarizers. {\bf (d)} Polarization image of a colloidal
membrane in which rods lie in the plane of the image. Directions
of polarizer and analyzer are indicated by white arrows. Scale
bars indicate 5 $\mu$m. }
\end{figure}

After a few hours, the {\it fd}/Dextran mixture prepared in region
2 completely phase separates with denser nematic tactoids
coalescing and settling to the bottom of the sample. In this case
a macroscopic I-N interface is formed. This makes it possible to
focus on the interface and directly observe the surface induced
smectic phase  (Fig.~\ref{figure3}). We conclude our description
of the system in region 2 by noting that there are no theoretical
predictions of the surface-induced smectic phase in rod/polymer
mixture. We expect that such phase is a result of non-monotonic
density profiles across the I-N interface~\cite{Shundyak02}.
Additionally, in the {\it fd}/polymer system rods in the surface
frozen layer lie in the plane of the interface. This is in
contrast to molecular systems which exhibit surface freezing where
anisotropic molecules are either tilted or perpendicular to the
interface~\cite{Wu93b,Ocko86}.

We now turn our attention to region 3 of the phase diagram. Right
after mixing the sample, in addition to the formation of nematic
droplets with a surface smectic, we observe self-assembly of rods
into disk-like or ribbon-like structures (Fig.~\ref{figure2}e and
f). The thickness of the disk corresponds to the length of a
single rod. When viewed from above a disk shows no birefringence
while from the side it shows maximum birefringence when oriented
at 45$^{o}$ with respect to the polarizer and analyzer
(Fig.~\ref{figure4}d). Therefore, polarization microscopy shows
that disks are composed of a monolayer of aligned rods in the
smectic-A configuration. We call these self-assembled disks
colloidal membranes because of their similarity to amphiphilic
membranes. Small homogeneously nucleated membranes
(Fig.~\ref{figure2}e) grow by coalescing laterally to form large
40 $\mu$ diameter isolated membranes
(Fig.~\ref{figure4}c)~\cite{Dogic01}. This suggests that an
isolated colloidal membrane and not a bulk smectic phase is the
equilibrium structure in region 3. Polarization microscopy
indicates that twisted ribbons are identical to disks except that
they have a twist along their long axis due to the chiral nature
of {\it fd}~\cite{Dogic00c}. We expect that the free energy
difference between these two morphologies is small and will
examine their relative stability elsewhere.

Real space images enables us to study the kinetic pathway for the
formation of colloidal membranes. They can either homogeneously
nucleate from the metastable isotropic suspension or can
heterogeneously nucleate at the surface-induced smectic phase
(Fig.~\ref{figure4}a and b). A colloidal membrane nucleated at the
interface grows into the isotropic phase either as a twisted
ribbon or a flat disk. Over a period of a few days twisted ribbons
can reach a lengths of several hundreds microns. Fluorescence
images indicate that there are no rods in the isotropic solution.
Therefore colloidal membranes (ribbons) must elongate due to rods
that diffuse from a metastable nematic phase through a surface
smectic to a more stable colloidal membrane. The fact that there
is a transport of rods across the interface shows that the
colloidal membranes are structures with lower free energy than the
nematic phase or bulk smectic phase. At lower degrees of
supercooling we mostly observe heterogenous surface induced
nucleation instead of homogeneous nucleation of colloidal
membranes. This shows that a two-step kinetic pathway has a lower
nucleation barrier for the formation of colloidal membranes. To
summarize, the phase separation in region 3 of the phase diagram
proceeds in two steps. In the first step on timescale of seconds
to minutes we observe the formation of nematic tactoids with
surface smectic phase identical to those observed in region 2.
However, these tactoids are metastable. In the second slow step on
a time scale of hours to months we observe the nucleation of
colloidal membranes at the surface frozen smectic phase and their
subsequent growth into the isotropic phase.

A few comments are in order regarding the structures observed in
region 3. First, to our knowledge this is the first time that
non-amhiphilic objects with very simple excluded volume
interactions have been self assembled into 2D membrane-like
(Fig.~\ref{figure4}c) and 1D polymer like structures
(Fig.~\ref{figure2}f)~\cite{Tkachenko02}. We speculate that these
structures are stabilized by protrusion like
fluctuations~\cite{Israelachvili91}. Second, it seems plausible
that isolated colloidal membranes observed in region 3 are highly
swollen lamellar phases previously observed in mixtures of nematic
{\it fd} and hard spheres~\cite{Adams98}. The swelling of the
lamellar phase is predicted theoretically, but has yet to be
observed in experiments~\cite{Koda96}. Third, as the osmotic
pressure is increased there is a transition to region 4 in which
small colloidal membranes irreversibly stack up on top of each
other to form elongated filaments (Fig.~\ref{figure2}g). The
nature of the transition from isolated membranes to a smectic
phase remains unexplored. Fourth, we observe metastable nematic
droplets wetted by a more stable surface induced smectic phase.
Usually the reverse effect is observed pathways where a stable
nucleus is wetted by a metastable
phase~\cite{tenWolde95,tenWolde97}

In conclusion, there are two important results that can be deduced
from our experiments. The first surprising result is that a
rod/polymer mixture exhibits surface freezing in which a quasi 2D
smectic phase wets the I-N interface. This effect occurs at rod
concentration of 235 mg/ml while bulk I-S phase transition occurs
at 255 mg/ml. To our knowledge this is the first time that the
surface freezing has been directly visualized in a system whose
phase behavior is dominated by entropic repulsive interactions.
The second result of this work is to demonstrate the relationship
between the surface freezing and the bulk isotropic-smectic phase
transition. A complex two step kinetic pathway for the nucleation
of the smectic phase out of the isotropic solution has been
identified. In the first step a metastable nematic droplet with a
surface frozen smectic phase nucleates in the isotropic solution.
In the next step isolated monolayers (colloidal membranes) of
smectic phase nucleate at the surface smectic phase and
subsequently grow into the isotropic phase. Due to the simplicity
and generality of the excluded volume interactions which dominate
the phase behavior of {\it fd}/Dextran mixture, the results
presented here should be relevant to a much wider class of systems
than those studied here.

I wish to thank Seth Fraden, Gerhard Gompper, Arjun Yodh, Tom
Lubensky, Daniel Chen, Peter Lang and Pavlik Lettinga for useful
discussions. I am particulary indebted to Jan Dhont for his
hospitality at FZ-Juelich and the Alexander von Humboldt
foundation for financial support. Part of this work was done at
Brandeis University where this research was supported by the
NSF-DMR grant to Seth Fraden.

\end{document}